\documentclass[osajnl,reprint]{revtex4-1}
\usepackage{graphicx}
\usepackage{subfigure}
\usepackage{amsmath,amssymb}
\usepackage[]{hyperref} 

\begin{abstract}
We present a semi-analytical formulation for calculating the supermodes and corresponding Bloch factors of light in hexagonal lattice photonic crystal waveguide arrays. We then use this formulation to easily calculate dispersion curves and predict propagation in systems too large to calculate using standard numerical methods.
\end{abstract}

\begin{document}
\title{Supermodes of Hexagonal Lattice Waveguide Arrays}
\author{J. Scott Brownless,$^{1,*}$ Felix J. Lawrence,$^{1}$ Sahand Mahmoodian,$^{1}$ Kokou B. Dossou,$^{2}$ Lindsay C. Botten,$^{2}$ and C. Martijn de Sterke$^{1}$}
\address{$^{1}$CUDOS and Institute of Photonics and Optical Science (IPOS), School of Physics, University of Sydney, NSW 2006, AUSTRALIA}
\address{$^{2}$CUDOS and Department of Mathematical Sciences, University of Technology, Sydney, NSW 2007, AUSTRALIA}
\address{$^*$Corresponding author: jbro@physics.usyd.edu.au}

\maketitle
\section{Introduction}
\label{intro}

Coupled optical waveguides are an important class of devices for the manipulation and processing of light and include, for example, the widely used directional coupler \cite{marcatili1969dielectric}. In waveguide arrays, which can be seen as generalizations of the directional coupler, the propagation of light is more complicated. The spreading out of light into neighboring waveguides in an array can be seen as a discrete version of diffraction. By varying the angle of the incoming light, Eisenberg \textit{et. al.} showed that the magnitude and sign of this discrete diffraction could be manipulated in ways not possible in regular diffraction \cite{eisenberg2000diffraction}. The modeling of directional couplers and complex waveguide arrays can be achieved with fully numerical methods, but the ability to predict unique phenomena such as discrete diffraction, is possible only by harnessing the efficiency and physical insight gained by modal methods.
Although an exact modal method exists, Yeh showed that much analytical insight can be achieved in the tight binding limit \cite{yeh1988optical}. In this limit the overarching coupled waveguide modes or supermodes of the system can be represented as linear combinations of the single waveguide modes with coupling between adjacent guides determined by the overlap between the modal fields and the neighboring waveguides. This method shows that for uniform waveguides the even superposition is always the fundamental coupled waveguide mode. This is consistent with the oscillation theorem, according to which, in a 1D geometry, the mode with the fewest number of nodes has the lowest energy \cite{ashcroft1976solid}. Yeh's method allows us to define the symmetry of the individual supermodes and accurately predicts the evolution of light in the individual waveguides within the array.

The coupling of photonic crystal waveguides (PCWs) has been a prominent area of research because of their short coupling lengths and unique dispersion properties, making them suitable for use in ultra compact devices. In square lattice PCW arrays, Locatelli \textit{et. al.} showed that, depending on the number of rows between the waveguides, the coupling coefficient can change sign, inverting the sign of the discrete diffraction coefficient throughout the array \cite{locatelli2006discrete}. The ability to invert the sign of the diffraction has applications in fields of negative index media and imaging. Locatelli's work arises from the work by de Sterke \textit{et. al.} into the coupled modes of square lattice PCWs  \cite{de2004modes}. They found that, as the number of rows between the waveguides varies, the symmetry of the fundamental mode alternates between even and odd because the underlying Bloch mode undergoes a $\pi$-phase change every transverse period. Therefore the order of the modes can invert, while still conforming to the oscillation theorem. Using a novel method combining the scattering matrix and tight binding methods, Botten \textit{et. al.} generalized this formulation to arrays with arbitrary numbers of waveguides \cite{botten2006tight}. They found that, as the number of rows between the waveguides is increased, the order of the supermodes completely inverts, which implies that the diffraction coefficient has reversed. In that work a single propagating plane wave approximation is made, mapping the problem onto that of the homogeneous waveguide array. Through node counting it is shown that the order of supermodes in the square lattice PCW array upholds the oscillation theorem.

\begin{figure}[hb]
	\centering
	\includegraphics[width=8cm]{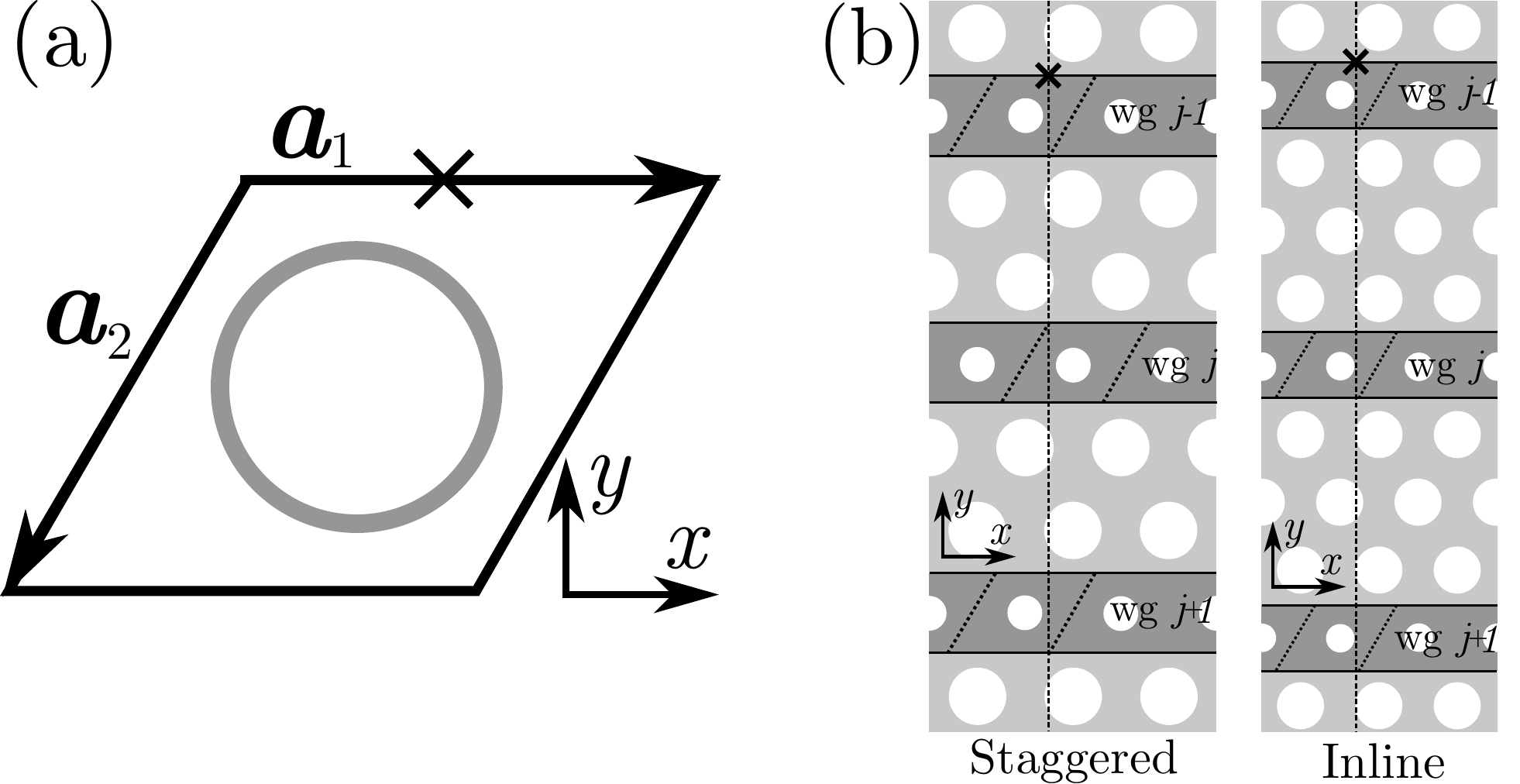}

	\caption{(a) Unit cell for a hexagonal lattice PC and corresponding lattice vectors. (b) The two different coupled waveguide geometries in a hexagonal lattice. Staggered (left), separated by an even number of defect rows, and in-line (right), separated by an odd number of defect rows.}\label{scat1}
\end{figure}

Despite the significant advantages of hexagonal lattice PCW systems, leading to their prominence in experimental research, modal methods for hexagonal lattice PCW systems are lagging behind their square lattice counterparts. This is because of the analytic complexities which arise when modeling the hexagonal lattice symmetry, for example, the non orthogonality of the lattice vectors (see Fig.~\ref{scat1}(a)).

In hexagonal lattice coupled PCWs, the geometry of the system can either be \textit{in-line} or \textit{staggered} depending on the number of rows between the waveguides (see Fig.~\ref{scat1}(b)). Ha \textit{et al.} have shown that this staggered geometry is particularly efficient in the coupling of slow light between the waveguides \cite{ha2008dispersionless}. In the staggered geometry, the coupled PCW modes are degenerate at the Brillouin zone (BZ) edge, where the two approaching modes have equal and opposite group velocities. Sukhorukov \textit{et. al.} showed that this is associated with the presence of vortex modes, satisfying the Bragg condition \cite{sukhorukov2009slow}. In previous work, we show that additional degeneracies can occur within the BZ, where the coupled PCW modes braid around each other. By analyzing the modes of the cladding, we find a region in the BZ where the two least evanescent Bloch modes decay at the same rate, the difference in the Bloch mode periodicity leads to the beating in field, causing braiding \cite{brownless2010coupled}.

Generalizing this theory to accommodate an array of coupled PCWs is not straightforward. There are a number of extra obstacles that arise in modeling this system. First, the advantage of using modal method analysis for complex problems, such as an array, comes from exploiting the symmetries of the system, but in the array these symmetries are much harder to find. To achieve the required symmetry in a system of more than two waveguides, we need to represent the field in each waveguide by one quantity which must symmetric about the waveguide center. But in this formulation, we wish to model systems where waveguides are created by modifying the radius or refractive index of a row of inclusions, rather than being completely removed. This means that we cannot treat the waveguide as a uniform medium---instead we must work completely in Bloch mode bases, in which we cannot easily relate the field at the center of the waveguide to its edges. Secondly, when comparing vertically aligned waveguide centers in the staggered geometry, we find they are at inequivalent points of the unit cell and therefore are not related by Bloch's theorem. Thirdly, since it has been shown there are two equally important Bloch modes in the hexagonal lattice \cite{brownless2010coupled}, the treatment must be generalized to the vector case, necessitating a treatment using matrices rather than scalars. Lastly, the basis vectors are not orthogonal which means that the Bloch factors in directions parallel and transverse to the waveguide do not decouple.

 For this reason we report here a formulation which allows us to calculate the modes of a arbitrary number of coupled PCWs in a hexagonal lattice. We consider the tight-binding limit, and, in spite of the complex geometry, find that the resultant matrix has the same symmetry as that of Yeh. The key difference is that the coupling between adjacent waveguides is expressed in terms of reflection and transmission matrices \cite{botten2006tight}, rather than overlap integrals. This allows for a simple, numerically efficient, tractable modeling tool which provides the physical understanding to exploit the properties of coupled hexagonal PCWs. We show, for example, how the braiding behavior, previously shown for two coupled PCWs, generalizes to the PCW array.

The outline of this paper is as follows: Section \ref{formulate} presents a detailed derivation of the formulation, showing the main result of this paper, before Section \ref{results} shows results achieved using the new formulation, comparing them to established methods and extending it to consider problems too complex to be practically calculated using standard numerical methods. Section \ref{conclusion} sums up the results of this paper.

\section{Modal Formulation}
\label{formulate}
In this section we derive a formulation for calculating the supermodes of a hexagonal lattice array with an arbitrary number of PC waveguides.
We begin by deriving a resonance condition that may be used to find the dispersion relation and modes of a single PCW.
We then calculate the reflection and transmission properties of the barriers between waveguides in the PCW array.
Finally we use these quantities, together with a tight binding approximation, to relate the fields in each waveguide and derive a resonance condition that may be used to find the dispersion relation and supermodes of the entire PCW array.

We work exclusively in the Bloch mode bases of the two PCs that comprise the system of coupled waveguides; the Bloch mode basis is the natural basis for a periodic medium.
At a single frequency, the field in a PC may be written as a superposition of propagating and evanescent Bloch modes \cite{botten2004bloch}, therefore we represent field by vectors of Bloch mode amplitudes.
In any given unit cell we may write a vector $\mathbf{c}^-$ of downward propagating/decaying mode amplitudes, and a vector $\mathbf{c}^+$ of upward propagating/decaying modes.

\begin{figure}
\centering
\includegraphics[width=8cm]{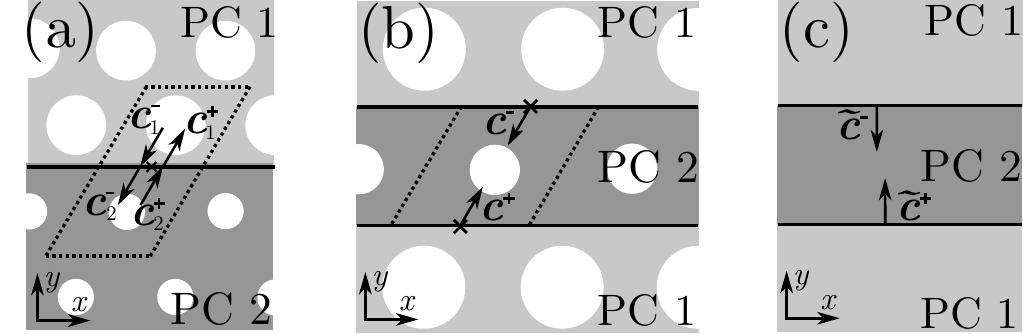}
\caption{(a) Bloch mode amplitudes defined across an interface between PC~1 and PC~2, described in Eq.~(\ref{interf}). (b) Bloch mode amplitudes defined in a PCW comprised on PC~2 sandwiched between layers of PC~1. (c) Shifted Bloch mode amplitudes defined in the same PCW as (b).}\label{figs1}
\end{figure}

In each bulk PC, the Bloch amplitudes in different unit cells are related by Bloch's theorem.
In our notation, the first lattice vector $\mathbf{a}_1=(d_x,0)$ is parallel to the waveguide.
The other lattice vector is $\mathbf{a}_2=(-d_x/2,-d_y)$ (see Fig.~\ref{scat1}(a)).
Bloch's theorem may be used to define the Bloch factor, $\mu_{Rs}=e^{i\mathbf{k}_s\cdot\mathbf{a}_R}$, which is the ratio of the amplitudes of mode $s$ at two points separated by the lattice vector $\mathbf{a}_R$.
Here $\mathbf{k}_s$ is the Bloch mode's wavevector; in our method, its $x$ component $k_x$ is specified for a particular frequency, and its $y$ component is found numerically using a Rayleigh multipole method \cite{botten2000formulation}.

Due to an impedance mismatch at the PC's edges, the Bloch modes may couple at an interface between PCs \cite{lawrence2009impedance}.
Therefore in our calculations we need to consider a vector of several Bloch modes in each PC.
The number of modes required for an accurate calculation varies and is related to the number of propagating grating diffraction orders; for the examples studied here, two Bloch modes in each direction are sufficient, but we include five modes for additional accuracy.
The unique mode-braiding behavior of hexagonal lattice PCWs arises from the existence of equally dominant evanescent Bloch modes \cite{brownless2010coupled}, and therefore fundamentally requires multiple Bloch modes to be modeled.

We can consider a PC as a stack of gratings and apply the grating diffraction equation.
At an interface between PCs where the periodicity in $x$ is conserved, coupling can only occur between modes with the same $k_x$
This means that for an entire system of modes, $\mu_{1s} = e^{i k_x d_x}$, which represents translation by $\mathbf{a}_1$, is fixed.
On the other hand, $\mu_{2s}$ is different for each mode.
Given a vector $\mathbf{c}^-$ of downward propagating/decaying mode amplitudes in a unit cell, the amplitudes in the unit cell positioned by $\mathbf{a}_2$ with respect to the first is $\mathbf{\Lambda} \mathbf{c}^-$, where $\mathbf{\Lambda} = \text{diag}(\mu_{2s})$.
For a vector $\mathbf{c}^+$ of upward propagating/decaying modes, the corresponding vector in the unit cell displaced by $\mathbf{a}_2$ is $\mathbf{\Lambda}' \mathbf{c}^+$, where $\mathbf{\Lambda}' = \text{diag}(\mu_{2s'})$.
It may be shown using reciprocity that for hexagonal lattice PC $\mathbf{\Lambda}'^{-1} = \mathbf{\Lambda}e^{ik_xd_x}$ \cite{botten2004bloch}.
When we relate Bloch mode amplitude vectors $\mathbf{c}^\pm$ in different unit cells, we generally do so for cells displaced by integer multiples of $\mathbf{a}_2$.

Each PC has its own set of Bloch modes. Across an interface between PCs (Fig.~\ref{figs1}(a)), the Bloch modes of the two PCs couple.
Analogously to thin film optics, the outgoing Bloch mode amplitudes $\mathbf{c}_1^+$ and $\mathbf{c}_2^-$ are related to the incoming Bloch mode amplitudes $\mathbf{c}_1^-$ and $\mathbf{c}_2^+$ by reflection and transmission matrices, which may be calculated most conveniently from numerically calculated PC impedances \cite{lawrence2009impedance}:
\begin{subequations}
\begin{equation}
	\mathbf{c}_1^+ = \mathbf{R}_{12} \mathbf{c}_1^- + \mathbf{T}_{21} \mathbf{c}_2^+,
\end{equation}
\begin{equation}
	\mathbf{c}_2^- = \mathbf{R}_{21} \mathbf{c}_2^+ + \mathbf{T}_{12} \mathbf{c}_1^-.
\end{equation}
\label{interf}
\end{subequations}

\subsection{The single waveguide problem}
\label{singlewgsec}
First, we derive the resonance condition for a single PC waveguide, which may be used to find the PCW's dispersion relation and modes.
This provides an introduction to our nomenclature and methods, as well as deriving a result we use in Sec.~\ref{sub:tight_binding} to find PCW array supermodes in which the field in the individual PCWs is approximately even.
The PC waveguide consists of a central waveguiding region (PC~2) sandwiched between two PC mirrors (PC~1).
The resonance condition is derived as a matrix analog to that of an optical slab waveguide.
We do not assume that the waveguiding region is a uniform dielectric: we allow the possibility that it consists of a row of holes---this allows an extra degree of freedom when tailoring the dispersion of the PCW.

We write down the relationship between the Bloch mode amplitudes on each side of the waveguide:
\begin{subequations}
	\label{eq:single_waveguides_a2}
	\begin{equation}
		\mathbf{c}^+ = \mathbf{R}_{21} \mathbf{\Lambda}_2 \mathbf{c}^-
	\end{equation}
	\begin{equation}
		\mathbf{c}^- = e^{ik_xd_x}\mathbf{R}_{21}' \mathbf{\Lambda}_2 \mathbf{c}^+,
	\end{equation}
\end{subequations}
where $\mathbf{c}^-$ is the vector of downward mode amplitudes at the top of the waveguide, and $\mathbf{c}^+$ is the vector of upward mode amplitudes at the bottom of the waveguide (Fig. \ref{figs1}b).
From the up-down symmetry surrounding the interface, we know $\mathbf{R}_{21}' = \mathbf{R}_{21}$.

The asymmetric factor $e^{ik_xd_x}$ arises due to the non-orthogonality of the two lattice vectors.
Its presence becomes particularly inconvenient in Sec.~\ref{sub:tight_binding}, which requires a high degree of symmetry.
We remove this factor by defining
\begin{equation}
	\label{eq:tilde_transformation}
	\mathbf{\tilde c}^{\pm} = \mathbf{c}^{\pm} e^{-ik_xx_{ROW}},
\end{equation}
where $x_{ROW}=(y_R/d_y)(d_x/2)$. The term $x_{ROW}$ corresponds to the shift in $x$ associated with a translation by $(-y_R/d_y)\mathbf{a}_2$, where $y_R$ is the $y$ coordinate of the unit cell for which $\mathbf{c}^\pm$ is defined.
Note that the $\mathbf{c}^+$ and $\mathbf{c}^-$ in Eqs. \eqref{eq:single_waveguides_a2} are defined on opposite sides of the waveguide region (Fig.~\ref{figs1}(b)), and so their $x_{ROW}$ differ by $d_x/2$: this is what removes the symmetry-breaking quantity $e^{ik_xd_x}$.
The original quantities $\mathbf{c}^\pm$ are defined in unit cells separated by multiples of the diagonal lattice vector $\mathbf{a}_2$.
For every second row of unit cells, multiplication by $e^{-ik_xx}$ corresponds to a translation by an integer multiple of $\mathbf{a}_1$, and so $\mathbf{\tilde c^\pm}$ has the straightforward interpretation of being the amplitude vector of the Bloch modes in the unit cell with $x=0$ in that row.
For the other rows, $\mathbf{\tilde c^\pm}$ has no direct interpretation, but $\mathbf{\tilde c^\pm}e^{ik_xd_x/2}$ is the vector of Bloch mode amplitudes at $x = d_x/2$.
This new basis for Bloch mode amplitudes necessitates the definition
\begin{equation}
	\label{eq:lambda_tilde_to_lambda}
	\mathbf{\tilde \Lambda} = \mathbf{\Lambda} e^{i k_x d_x/2},
\end{equation}
where $\mathbf{\tilde \Lambda}$ replaces $\mathbf{\Lambda}$ in relating the downwards amplitudes of Bloch modes in different rows of a PC.
The reflection and transmission matrices are unchanged.

With our new quantities, we can rewrite Eqs.~\eqref{eq:single_waveguides_a2} as
\begin{subequations}
	\label{eq:swgrels}
	\begin{align}
	\label{swgrels1}
	 \mathbf{\tilde{c}}^+&=\mathbf{R}_{21}\mathbf{\tilde{\Lambda}}_{2}\mathbf{\tilde{c}}^-,\\
	 \mathbf{\tilde{c}}^-&=\mathbf{R}_{21}\mathbf{\tilde{\Lambda}}_{2}\mathbf{\tilde{c}}^+,
	\label{swgrels2}
	\end{align}
\end{subequations}
relating shifted Bloch mode amplitudes (Fig.~\ref{figs1}(c))
By substituting Eq.~(\ref{swgrels2}) into Eq.~(\ref{swgrels1}), we find the resonance condition for a single PCW mode,
\begin{align}
\label{swgres}
\mathbf{\tilde{c}}^-&=\mathbf{R}_{21}\mathbf{\tilde{\Lambda}}_{2}\mathbf{R}_{21}\mathbf{\tilde{\Lambda}}_{2}\mathbf{\tilde{c}}^-,\\
\mathbf{0}&=(\mathbf{I}\pm\mathbf{R}_{21}\mathbf{\tilde{\Lambda}}_{2})\mathbf{\tilde{c}}^-.\label{eq:swgres_factorized}
\end{align}
So, the condition for a single waveguide mode is
\begin{equation}
	\text{det}(\mathbf{I}\pm\mathbf{R}_{21}\mathbf{\tilde{\Lambda}}_{2}) = 0.
\end{equation}
The eigenvector $\mathbf{\tilde{c}}^-$ associated with the zero eigenvalue gives the waveguide mode in terms of the Bloch modes of PC~2, the in-band PC.
Using Eqs.~\eqref{eq:swgrels}~and~\eqref{eq:swgres_factorized}, we derive
\begin{subequations}
	\begin{align}
		\mathbf{\tilde{c}}^+ &= +\mathbf{\tilde{c}}^-,\text{ or} \label{eq:evres}\\
		\mathbf{\tilde{c}}^+ &= -\mathbf{\tilde{c}}^-, \label{eq:oddres}
	\end{align}
\end{subequations}
where Eq.~\eqref{eq:evres} holds for even PCW modes and Eq.~\eqref{eq:oddres} holds for odd PCW modes.
In an experiment it is easier to couple light into symmetric modes, so we will concentrate on the even PCW modes.

The contribution of even modes to the field in a waveguide may be summarized as
\begin{align}
	\mathbf{\tilde{e}} = \frac{1}{2}(\mathbf{\tilde{c}}^-+\mathbf{\tilde{c}}^+), \label{eq:e_def_sw}
\end{align}
which may be shown to represent the amplitudes of the even superpositions of Bloch modes in a single waveguide (Eq.~\eqref{swgfield2}).

To find the waveguide mode's field $H(x,y)$ inside a waveguide unit cell, we convert our calculated result, $\mathbf{\tilde c}^+$ and $\mathbf{\tilde c}^-$, back into the basis of $\mathbf{c}^+$ and $\mathbf{c}^-$.
If we choose $\mathbf{c}^-$ such that it is defined with respect to the origin, then by Eq.~\eqref{eq:tilde_transformation} there is no difference between the two bases, i.e., $\mathbf{c}^- = \mathbf{\tilde c}^-$ and the amplitudes of the upward Bloch modes in this cell are $\mathbf{\tilde{\Lambda}}_2\mathbf{\tilde{c}}^+$.
Therefore the magnetic field $H_{swg}$ at position $\mathbf{r}$ in the unit cell is
\begin{align}
\label{swgfield}
H_{{\rm swg}}(\mathbf{r}) &= (\mathbf{\tilde{\Lambda}}_2\mathbf{\tilde{c}}^+)^{T}\mathbf{\Psi}^+(\mathbf{r}) + \mathbf{\tilde{c}}^{-T}\mathbf{\Psi}^-(\mathbf{r}),\\
&= \mathbf{\tilde{e}}^{T}(\mathbf{\tilde{\Lambda}}_2 \mathbf{\Psi}^+(\mathbf{r}) + \mathbf{\Psi}^-(\mathbf{r}))\label{swgfield2}
\end{align}
where $\mathbf{\Psi}^+(\mathbf{r})$ and $\mathbf{\Psi}^-(\mathbf{r})$ are respectively vectors of the fields of the forward and backward Bloch modes.
A similar expression may be written for the electric field.
It is straightforward to use the transmission matrix $\mathbf{T}_{21}$ to calculate the Bloch mode amplitudes of PC~1 from $\mathbf{\tilde c}^+$, and a similar procedure may be followed to calculate the field of the waveguide mode throughout PC~1.

\subsection{The transmission and reflection matrices across the waveguide barrier}
In this paper we consider PCW arrays consisting of layers of two different photonic crystals, PC~1 and PC~2, where PC~1 is the bulk PC, which is in bandgap, and PC~2 is the waveguide .
The barrier consists of $\ell$ rows of PC~1, and the waveguide is one row thick.
PC~1 and PC~2 may be dielectrics or PCs, and need not have the same background refractive index.
\begin{figure}
\centering
\includegraphics[width=8cm]{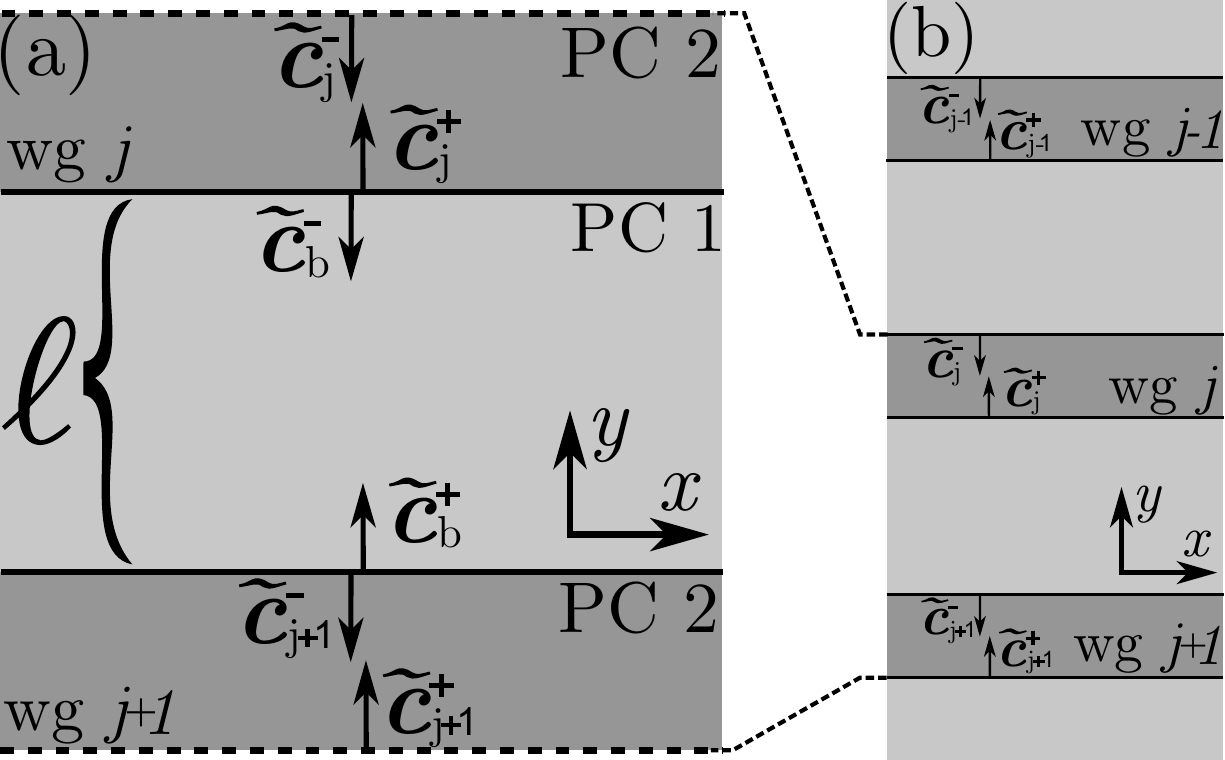}
\caption{(a) Bloch mode amplitudes defined across a barrier consisting of $\ell$ rows of PC~1 between two waveguides. The broken lines indicate a perfectly matched layer, rather than an interface.(b) Bloch mode amplitudes defined in a section of the PCW array.}\label{figs2}
\end{figure}

This section focuses on a part of the array consisting of just two neighboring waveguides and the barrier separating them (see Fig.~\ref{figs2}a), to find the reflection and transmission matrices across the barrier.
We neglect reflections off the outermost edge of each waveguide, and aim to relate the vectors $\mathbf{\tilde c}_j^-$, $\mathbf{\tilde c}_j^+$, $\mathbf{\tilde c}_{j+1}^-$ and $\mathbf{\tilde c}_{j+1}^+$ to find the reflection and transmission matrices in each direction across the barrier,
\begin{subequations}
\begin{align}
\label{barrel1}
\mathbf{\tilde{c}}_{j}^+&=\mathbf{\tilde{R}}\mathbf{\tilde{\Lambda}}_2\mathbf{\tilde{c}}_{j}^-+\mathbf{\tilde{T}}'\mathbf{\tilde{\Lambda}}_2\mathbf{\tilde{c}}_{j+1}^+,\\
\label{barrel2}
\mathbf{\tilde{c}}_{j+1}^-&=\mathbf{\tilde{T}}\mathbf{\tilde{\Lambda}}_2\mathbf{\tilde{c}}_{j}^-+\mathbf{\tilde{R}}'\mathbf{\tilde{\Lambda}}_2\mathbf{\tilde{c}}_{j+1}^+,
\end{align}
\end{subequations}
where $\mathbf{\tilde{R}}, \mathbf{\tilde{T}}$ and $\mathbf{\tilde{R}}',\mathbf{\tilde{T}}'$ are reflection and transmission matrices from above and below the barrier respectively.

If we set $\mathbf{\tilde{c}}_{j+1}^+=\mathbf{0}$ (see Fig.~\ref{figs2}a), then the only Bloch mode incident on the barrier is from above, i.e. $\mathbf{\tilde{c}}_{j}^-$, and our relations in Eqs.~(\ref{barrel1}) and~(\ref{barrel2}) become $\mathbf{\tilde{c}}_{j}^+=\mathbf{R}\mathbf{\tilde{c}}_{j}^-$ and $\mathbf{\tilde{c}}_{j+1}^-=\mathbf{T}\mathbf{\tilde{c}}_{j}^-$.
The relations between $\mathbf{\tilde{c}}_{j}^+$ and $\mathbf{\tilde{c}}_{j+1}^-$ and the Bloch modes inside the barrier, $\mathbf{\tilde{c}}_b^\pm$ are given by
\begin{subequations}
	\begin{align}
	 \mathbf{\tilde{c}}_{j}^+&=\mathbf{R}_{21}\mathbf{\tilde{\Lambda}}_2\mathbf{\tilde{c}}_{j}^-+\mathbf{T}'_{12}\mathbf{\tilde{\Lambda}}^{\ell}_1\mathbf{\tilde{c}}_b^+,\\
	 \mathbf{\tilde{c}}_{j+1}^-&=\mathbf{T}_{12}\mathbf{\tilde{\Lambda}}^{\ell}_1\mathbf{\tilde{c}}_b^-,\\
	 \mathbf{\tilde{c}}_b^+&=\mathbf{R}_{12}\mathbf{\tilde{\Lambda}}_1^\ell\mathbf{\tilde{c}}_b^-,\\
	 \mathbf{\tilde{c}}_b^-&=\mathbf{R}'_{12}\mathbf{\tilde{\Lambda}}_1^{\ell}\mathbf{\tilde{c}}_b^++\mathbf{T}_{21}\mathbf{\tilde{\Lambda}}_2\mathbf{\tilde{c}}_{j+1}^+.
	\end{align}
\end{subequations}
From the up-down symmetry, we know $\mathbf{T}_{12}' = \mathbf{T}_{12}$ and $\mathbf{R}_{12}' = \mathbf{R}_{12}$.
We can rearrange these relations to find $\mathbf{\tilde{c}}_{j+1}^-$ in terms of $\mathbf{\tilde{\Lambda}}_2\mathbf{\tilde{c}}_{j}^-$:
\begin{align}
\mathbf{\tilde{c}}_{j+1}^-=&\mathbf{T}_{12} \mathbf{\tilde{\Lambda}}_{1}^\ell(\mathbf{I}-\mathbf{R}_{12}\mathbf{\tilde{\Lambda}}_{1}^{\ell}\mathbf{R}_{12} \mathbf{\tilde{\Lambda}}_{1}^\ell)^{-1} \mathbf{T}_{21}\mathbf{\tilde{\Lambda}}_2\mathbf{\tilde{c}}_{j}^-=\mathbf{\tilde{T}}\mathbf{\tilde{\Lambda}}_2\mathbf{\tilde{c}}_{j}^-,\\
\label{eq:barrier_transmission}
\Rightarrow\mathbf{\tilde{T}}=&\mathbf{T}_{12} \mathbf{\tilde{\Lambda}}_{1}^\ell(\mathbf{I}-\mathbf{R}_{12}\mathbf{\tilde{\Lambda}}_{1}^{\ell}\mathbf{R}_{12} \mathbf{\tilde{\Lambda}}_{1}^\ell)^{-1} \mathbf{T}_{21}.
\end{align}
Similarly, by rearranging the relations to find $\mathbf{\tilde{c}}_{j}^+$ in terms of $\mathbf{\tilde{\Lambda}}_2\mathbf{\tilde{c}}_{j}^-$, we derive
\begin{equation}
\mathbf{\tilde{R}}=\mathbf{R}_{21}+\mathbf{T}_{12}\mathbf{\tilde{\Lambda}}_{1}^\ell\mathbf{R}_{12}\mathbf{\tilde{\Lambda}}_{1}^\ell(\mathbf{I}-\mathbf{R}_{12}\mathbf{\tilde{\Lambda}}_{1}^\ell \mathbf{R}_{12}\mathbf{\tilde{\Lambda}}_{1}^\ell)^{-1}\mathbf{T}_{21}.
\label{eq:barrier_reflection}
\end{equation}

If we instead set $\mathbf{\tilde{c}}_{j}^-=\mathbf{0}$, then Eqs.~(\ref{barrel1}) and~(\ref{barrel2}) become $\mathbf{\tilde{c}}_{j}^+=\mathbf{T}'\mathbf{\tilde{c}}_{j+1}^+$ and $\mathbf{\tilde{c}}_{j+1}^-=\mathbf{R}'\mathbf{\tilde{c}}_{j+1}^+$. By following the same method as above, we derive $\mathbf{\tilde T}' = \mathbf{\tilde T}$ and $\mathbf{\tilde R}' = \mathbf{\tilde R}$.

\subsection{PCW array dispersion relation and supermodes}
\label{sub:tight_binding}
In a waveguide array, interactions occur between every pair of waveguides.
For large arrays of many waveguides, it is unfeasible to model all these interactions.
Following Yeh \cite{yeh1988optical}, we ignore the interactions between non-adjacent waveguides, making a \textit{nearest neighbor} approximation, which is valid in the tight binding regime.
In doing so, we assume that the field decays substantially across the barriers between waveguides, i.e., we assume that $|\mu^\ell| \ll 1$ for every element $\mu$ of $\mathbf{\tilde{\Lambda}}_1$.
Formally, we neglect terms of order $O(\mathbf{\tilde{\Lambda}}_1^{2\ell})$.
Doing so excludes the interactions between non-adjacent waveguides, as well as the Fabry-Perot-like terms in the interactions between adjacent waveguides.

Under this approximation, the transmission and reflection matrices in Eqs.~(\ref{eq:barrier_transmission}) and~(\ref{eq:barrier_reflection}) simplify to
\begin{subequations}
	\begin{equation}
		\mathbf{\tilde{T}} = \mathbf{T}_{12} \mathbf{\tilde{\Lambda}}_{1}^\ell \mathbf{T}_{21},
	\end{equation}
	\begin{equation}
		\mathbf{\tilde{R}}=\mathbf{R}_{21}.
	\end{equation}
\end{subequations}
The reflection matrix for the waveguide barrier is approximated by the reflection off an infinitely thick barrier since any effect from the neighboring waveguide consists of terms of order $O(\mathbf{\tilde{\Lambda}}_1^{2\ell})$.

We can now relate the Bloch mode amplitudes of each of the waveguides in an $M$-waveguide PCW array.
The vectors $\tilde{\mathbf{c}}_j^+$ and $\tilde{\mathbf{c}}_j^-$ that represent upward and downward Bloch mode amplitudes in waveguide $j$ are as defined in the previous section.
The topmost and bottommost waveguides, 1 and $M$, interact with only one neighbor each.
The relationship between $\mathbf{\tilde{c}}_1^-$ and $\mathbf{\tilde{c}}_1^+$ is similar to that of a standalone PCW (Eq.~\eqref{eq:swgrels}); the difference is that $\mathbf{\tilde{c}}_1^+$ includes a contribution $\mathbf{\tilde{T}}\mathbf{\tilde{\Lambda}}_{2}\mathbf{\tilde{c}}_{2}^+$ from waveguide 2:
\begin{subequations}
	\label{eq:coupledrelswg1}
	\begin{align}
		\label{coupledrels1}
		\mathbf{\tilde{c}}_1^-&= \mathbf{R}_{21}\mathbf{\tilde{\Lambda}}_{2}\mathbf{\tilde{c}}_1^+,\\
		\label{coupledrels2}
		\mathbf{\tilde{c}}_1^+&= \mathbf{R}_{21}\mathbf{\tilde{\Lambda}}_{2}\mathbf{\tilde{c}}_1^- + \mathbf{\tilde{T}}\mathbf{\tilde{\Lambda}}_{2}\mathbf{\tilde{c}}_{2}^+.
	\end{align}
\end{subequations}
Similarly, in waveguide $M$, $\mathbf{\tilde{c}}_{M}^-$ includes a contribution from waveguide $M-1$:
\begin{subequations}
	\label{eq:coupledrelswgM}
	\begin{align}
	\label{coupledrels5}
	 \mathbf{\tilde{c}}_{M}^-&=\mathbf{R}_{21}\mathbf{\tilde{\Lambda}}_{2}\mathbf{\tilde{c}}_M^++\mathbf{\tilde{T}}\mathbf{\tilde{\Lambda}}_{2}\mathbf{\tilde{c}}_{M-1}^-,\\
	\label{coupledrels6}
	 \mathbf{\tilde{c}}_{M}^+&=\mathbf{R}_{21}\mathbf{\tilde{\Lambda}}_{2}\mathbf{\tilde{c}}_M^-.
	\end{align}
\end{subequations}
The other waveguides $1<j<M$, interact with both their neighbors (see Fig.~\ref{figs2}b):
\begin{subequations}
	\label{eq:coupledrelswgj}
	\begin{align}
	\label{coupledrels3}
	 \mathbf{\tilde{c}}_j^-&=\mathbf{R}_{21}\mathbf{\tilde{\Lambda}}_{2}\mathbf{\tilde{c}}_j^++\mathbf{\tilde{T}}\mathbf{\tilde{\Lambda}}_{2}\mathbf{\tilde{c}}_{j-1}^-,\\
	\label{coupledrels4}
	 \mathbf{\tilde{c}}_j^+&=\mathbf{R}_{21}\mathbf{\tilde{\Lambda}}_{2}\mathbf{\tilde{c}}_j^-+\mathbf{\tilde{T}}\mathbf{\tilde{\Lambda}}_{2}\mathbf{\tilde{c}}_{j+1}^+.
	\end{align}
\end{subequations}

In these equations, we represent the field in each waveguide by two different quantities: on the top edge as the amplitude vector of the forward modes $\mathbf{\tilde{c}}^-$, then on the bottom edge as the backward modes $\mathbf{\tilde{c}}^+$.
In order to apply methods analogous to those used with coupled dielectric waveguides \cite{yeh1988optical}, we need to represent the field in each waveguide by one quantity.
We define
\begin{equation}
	\label{edefine}
	 \mathbf{\tilde{e}}_j=\frac{1}{2}(\mathbf{\tilde{c}}_j^-+\mathbf{\tilde{c}}_j^+).
\end{equation}
$\mathbf{\tilde{e}}_j$ only represents the contributions of even superpositions of Bloch modes in the waveguide, as noted in Sec.~\ref{singlewgsec}.
From here on we represent the amplitudes in waveguide $j$ by $\mathbf{\tilde{e}}_j$.
We start by using Eq.~\eqref{edefine} to write the $2M$ equations~\eqref{eq:coupledrelswg1}--\eqref{eq:coupledrelswgj} as $M$ equations relating the $\mathbf{\tilde{e}}_j$:
\begin{subequations}
	\label{eq:coupledrelse}
	\begin{align}
	\label{coupledrelse2}
	\mathbf{\tilde{e}}_1 &= \mathbf{\tilde{R}}\mathbf{\tilde{\Lambda}}_{2}\mathbf{\tilde{e}}_1 + \frac{1}{2}\mathbf{\tilde{T}}\mathbf{\tilde{\Lambda}}_{2}\mathbf{\tilde{c}}_{2}^+,\\
	\label{coupledrelse1}
	\mathbf{\tilde{e}}_j &= \mathbf{\tilde{R}}\mathbf{\tilde{\Lambda}}_{2}\mathbf{\tilde{e}}_j + \frac{1}{2}\mathbf{\tilde{T}}\mathbf{\tilde{\Lambda}}_{2}\mathbf{\tilde{c}}_{j+1}^+ + \frac{1}{2}\mathbf{\tilde{T}}\mathbf{\tilde{\Lambda}}_{2}\mathbf{\tilde{c}}_{j-1}^-,\\
	\label{coupledrelse3}
	\mathbf{\tilde{e}}_M &= \mathbf{\tilde{R}}\mathbf{\tilde{\Lambda}}_{2}\mathbf{\tilde{e}}_M + \frac{1}{2}\mathbf{\tilde{T}}\mathbf{\tilde{\Lambda}}_{2}\mathbf{\tilde{c}}_{M-1}^-.
	\end{align}
\end{subequations}
In order to replace the residual $\mathbf{\tilde{c}}_j^\pm$ in the last terms of these equations, we recall our approximation discarding terms of order $O(\mathbf{\tilde{\Lambda}}_1^{2\ell})$.
The field in each waveguide is an even single waveguide mode, plus interactions of order $O(\mathbf{\tilde{\Lambda}}_1^{\ell})$ from adjacent waveguides: thus for each waveguide in the array, using Eq.~\eqref{eq:evres} we write $\tilde{\mathbf{c}}_j^+ = \tilde{\mathbf{c}}_j^- + O(\mathbf{\tilde{\Lambda}}_1^{\ell})$.
Note that the $\mathbf{\tilde{c}}_j^\pm$ terms in Eq.~\eqref{eq:coupledrelse} are all multiplied by $\mathbf{\tilde{T}}$, which is of order $O(\mathbf{\tilde{\Lambda}}_1^{\ell})$.
Therefore for these terms we can ignore the $O(\mathbf{\tilde{\Lambda}}_1^{\ell})$ contributions from other waveguides and write $\tilde{\mathbf{c}}_j^\pm \simeq \tilde{\mathbf{e}}_j$.

Applying this substitution to Eq.~\eqref{eq:coupledrelse}, we write the equations solely in terms of $\mathbf{\tilde{e}}_j$, in the form of a tri-diagonal block matrix equation,
\begin{equation}
\label{virtualmatrix}
\begin{pmatrix}
\mathbf{A} & \mathbf{B} & \mathbf{0} & \dotsi & \mathbf{0}\\
\mathbf{B} & \mathbf{A} & \mathbf{B} & \cdots & \mathbf{0}\\
\mathbf{0} & \mathbf{B} & \mathbf{A} & \cdots & \mathbf{0}\\
\vdots   	 & \vdots     & \vdots     & \ddots & \vdots    \\
\mathbf{0} & \mathbf{0} & \mathbf{0} & \cdots & \mathbf{A}&
\end{pmatrix}
\begin{pmatrix}
\mathbf{\tilde{e}}_1\\
\mathbf{\tilde{e}}_2\\
\mathbf{\tilde{e}}_3\\
\vdots\\
\mathbf{\tilde{e}}_M
\end{pmatrix}
=
\begin{pmatrix}
\mathbf{\tilde{e}}_1\\
\mathbf{\tilde{e}}_2\\
\mathbf{\tilde{e}}_3\\
\vdots\\
\mathbf{\tilde{e}}_M
\end{pmatrix},
\end{equation}
where $\mathbf{A}=\mathbf{R}_{21}\mathbf{\tilde{\Lambda}}_{2}$ and $\mathbf{B}=\frac{1}{2}\mathbf{\tilde{T}}\mathbf{\tilde{\Lambda}}_{2}$.

This is the resonance condition for a coupled waveguide array, in the same way that Eq.~(\ref{swgres}) is the resonance condition for a single waveguide.
For each frequency, a range of $k_x$ can be scanned for the existence of supermodes by testing whether the condition may be satisfied by each individual $k_x$.

The difference between Eq.~(\ref{virtualmatrix}), and the expression for conventional waveguide arrays \cite{yeh1988optical}, is that each element in Eq.~\eqref{virtualmatrix} is a $n \times n$ matrix rather than a scalar.
If $n$ Bloch modes are considered in each waveguide, then the matrix in Eq.~\eqref{virtualmatrix} is of dimension $nM$.
For a large number of waveguides $M$, it can be computationally expensive to solve the eigensystem of this matrix.
Therefore, in a block-matrix analog to the conventional treatment \cite{yeh1988optical}, we exploit the tridiagonal block symmetry of the matrix and block-diagonalize it analytically using a similarity transformation.
The problem then reduces to solving the eigensystems of $M$ matrices, each with dimension $n$.
Analogous to Yeh's treatment \cite{yeh1988optical}, the elements of the symmetric and unitary similarity transform $\mathbf{\Phi}$ that block-diagonalizes the tridiagonal matrix are given by
\begin{equation}
	\label{eq:Phiphi_def}
	\Phi_{js}=\phi_{js} \mathbf{I},
\end{equation}
 where $\mathbf{I}$ has dimension $n$ and
\begin{equation}
	\label{eq:Phi_def}
	\phi_{js} = \sqrt{\frac{2}{M+1}}\,\sin\left(\frac{js\pi}{M+1}\right).
\end{equation}

To demonstrate how to use Eq.~\eqref{virtualmatrix} to calculate a dispersion relation, we give the example of the two waveguide case, for which
\begin{equation}
	\label{virtualmatrix3}
	\begin{pmatrix}
		\mathbf{A}& \mathbf{B}\\
		\mathbf{B}&\mathbf{A}\\
	\end{pmatrix}
	\begin{pmatrix}
		\mathbf{\tilde{e}}_1\\ \mathbf{\tilde{e}}_2\\
	\end{pmatrix}= \begin{pmatrix}
	\mathbf{\tilde{e}}_1\\ \mathbf{\tilde{e}}_2\\
	\end{pmatrix}.
\end{equation}
This can be block diagonalized using a similarity transformation to give
\begin{equation}
	\label{virtualmatrix3o}
	\mathbf{\Phi}^{-1}\begin{pmatrix}
	\mathbf{A}& \mathbf{B}\\
	\mathbf{B}&\mathbf{A}\\
\end{pmatrix}
\mathbf{\Phi}\mathbf{\Phi}^{-1}
\begin{pmatrix}
	\mathbf{\tilde{e}}_1\\ \mathbf{\tilde{e}}_2\\
\end{pmatrix}= \mathbf{\Phi}^{-1}\begin{pmatrix}
\mathbf{\tilde{e}}_1\\ \mathbf{\tilde{e}}_2\\
\end{pmatrix},
\end{equation}
where, from Eq.~(\ref{eq:Phi_def}) $\mathbf{\Phi}=\frac{1}{\sqrt{2}}\left( \begin{smallmatrix}
\mathbf{I}&\mathbf{I}\\ \mathbf{I}&-\mathbf{I}
\end{smallmatrix} \right)$, giving
\begin{equation}
	\label{virtualmatrix3diag}
	\begin{pmatrix}
		\mathbf{A+B}& \mathbf{0}\\
		\mathbf{0}&\mathbf{A-B}\\
	\end{pmatrix}
	\begin{pmatrix}
		\mathbf{\tilde{e}}_1+\mathbf{\tilde{e}}_2\\ \mathbf{\tilde{e}}_1-\mathbf{\tilde{e}}_2
	\end{pmatrix}=\begin{pmatrix}
	\mathbf{\tilde{e}}_1+\mathbf{\tilde{e}}_2\\ \mathbf{\tilde{e}}_1-\mathbf{\tilde{e}}_2
\end{pmatrix}\equiv\begin{pmatrix}
\mathbf{\tilde{x}}_1\\ \mathbf{\tilde{x}}_2
\end{pmatrix}.
\end{equation}
Equation~(\ref{virtualmatrix3diag}) decouples into two nontrivial solutions:
$(\mathbf{I}-(\mathbf{A+B})) \mathbf{\tilde{x}}_1 = \mathbf{0}$ with $\mathbf{\tilde{x}}_2 = \mathbf{0}$
(even supermode),
and
$(\mathbf{I}-(\mathbf{A-B})) \mathbf{\tilde{x}}_2 = \mathbf{0}$ with $\mathbf{\tilde{x}}_1 = \mathbf{0}$
(odd supermode).
The coupled waveguide array's dispersion relation may be constructed from the frequency-$k_x$ pairs at which at least one of these conditions is satisfied.
So the condition for an even supermode is $\text{det}(\mathbf{I}- (\mathbf{A+B})) = \mathbf{0}$, and for this mode $\mathbf{\tilde{e}}_1 = \mathbf{\tilde{e}}_2 =\frac{1}{2}\mathbf{\tilde{x}}_1$, where $\mathbf{\tilde{x}}_1$ is an eigenvector of $\mathbf{I}- (\mathbf{A+B})$ with an eigenvalue of zero.  This is the Bloch mode analogue of the result achieved in our previous work \cite{brownless2010coupled}.

Due to the block tridiagonal symmetry of the matrix in Eq.~\eqref{virtualmatrix}, similar conditions may be obtained even for very large systems of coupled waveguides.
Generally, the condition for the $s$th PCW array supermode is
\begin{equation}
(\mathbf{A}-\sigma_s\mathbf{B})\mathbf{\tilde{x}}_s=\mathbf{\tilde{x}}_s,
\label{virtualequations}
\end{equation}
with all other $\mathbf{\tilde{x}}_i = \mathbf{0}$, where $\sigma_s=2\cos(s\pi/(M+1))$. \cite{yeh1988optical}. The fields in each waveguide for supermode $s$ can be found by
\begin{equation}
	\label{eq:a_of_x}
	\begin{pmatrix}
		\mathbf{\tilde e}_1 \\
		\mathbf{\tilde e}_2 \\
		\vdots
	\end{pmatrix}
	=
	\mathbf{\Phi}_s\mathbf{x}_s,
\end{equation}
where $\mathbf{\Phi}_s$ denotes block column $s$ of $\mathbf{\Phi}$

The procedure to find the dispersion relation of a coupled PCW array therefore involves calculating $\mathbf{A}$ and $\mathbf{B}$ for frequency-$k_x$ pair and checking whether $\text{det}(\mathbf{I}-(\mathbf{A} - \sigma_s \mathbf{B})) = 0$ for any $\sigma_s$.  At a frequency and $k_x$ for which this determinant is zero, then the null vector of $\mathbf{I}-(\mathbf{A} - \sigma_s \mathbf{B})$ is the vector $\mathbf{\tilde x}_s$.

The resonance condition Eq.~\eqref{virtualequations} can be written directly in terms of the original reflection, transmission and propagation matrices, using Eq.~\eqref{eq:lambda_tilde_to_lambda}:
\begin{equation}
(\mathbf{I}-e^{ik_xd/2}\mathbf{R}_{21}\mathbf{\Lambda}_{2f}+\frac{\sigma_s}{2}e^{ik_xd(\ell+1)/2}\mathbf{T}_{12}\mathbf{\Lambda}_{1f}^\ell\mathbf{T}_{21}\mathbf{\Lambda}_{2f})\mathbf{\tilde{x}}_s=\mathbf{0}.
\label{finalcondition}
\end{equation}
The first two terms in Eq.~(\ref{finalcondition}) give the resonance condition of the single waveguide in a hexagonal lattice. Therefore the third term, of order $\mathbf{\Lambda}_1^\ell$, is the effect of the neighboring waveguides in the tight-binding approximation. In the photonic band gap, all Bloch modes in the cladding are evanescent and have Bloch factors $|\mu_1|<1$. As we increase the distance between waveguides, $\ell$ increases, and $\mathbf{\Lambda}_{1}^\ell\to\mathbf{0}$. Therefore the resonance condition of the array approaches that of the single waveguide as we increase the distance between the waveguides, as expected.

\subsection{PCW array modes and propagation} 
\label{sub:modes_and_propagation}
At the $k_x$ associated with supermode $s$, Eq.~\eqref{virtualequations} may be used to calculate $\mathbf{\tilde x}_s$.
The other $\mathbf{\tilde x}_i = \mathbf{0}$, and these values are used implicitly in Eq.~\eqref{eq:a_of_x} to find the supermode's field amplitudes $\mathbf{\tilde e}_j$ in each waveguide $j$.

In choosing to represent the field in each waveguide by the vector $\mathbf{\tilde e}_j$, we assumed that the field is an even superposition of upward and downward Bloch modes in each waveguide.
In Yeh's treatment of coupled waveguides \cite{yeh1988optical}, the field in the  waveguide array is written as a superposition of single waveguide modes, with one amplitude per waveguide.
We do the same by writing $\mathbf{\tilde x}_s \simeq \mathbf{\tilde e}$, where $\mathbf{\tilde e}$ is the vector of Bloch mode amplitudes calculated for a single waveguide.
This approximation is consistent with an eigenvalue perturbation theory approach: $k_x$ is found including terms of order $O(\mathbf{\tilde \Lambda}^\ell)$, but the modal field is found ignoring these terms.

From Eq.~\eqref{eq:a_of_x}, it follows that the field of supermode $s$ in waveguide $j$ is
\begin{equation}
	\label{eq:Phi_phi}
	\mathbf{\tilde e}_j = \Phi_{js} \mathbf{\tilde e} = \phi_{js} \mathbf{\tilde e}.
\end{equation}

We now generalize to consider a superposition of supermodes with amplitudes $w_s$, and calculate the resulting field in each waveguide, $\mathbf{\tilde e}_j =\tilde{a}_j\mathbf{\tilde{e}}$, which we simply represent by the scalar amplitude $\tilde{a}_j$. Then
\begin{subequations}
\begin{align}
\label{eq:initialsum}
\tilde{a}_j=&\sum_s{\phi_{js}w_s},\\
\label{eq:initialmat}
{\rm i.e. }~\mathbf{\tilde{a}}=&\mathbf{\phi}\mathbf{w},
\end{align}
\end{subequations}
where $\mathbf{\tilde a}$ and $\mathbf{w}$ are vectors of $\tilde a_j$ and $w_s$ respectively.

Propagation in the PCW array is most simply represented in the basis of its supermodes, each of which simply acquires phase as it propagates. We use the propagation constant $k_x$ of each supermode $s$ to write its Bloch factor $\mu_{1s}=\exp(i k_x d_x)$
Given an initial field $\mathbf{w}(0) = \phi^{-1} \mathbf{\tilde{a}}(0)$ at the start of the waveguide array, the amplitude in waveguide $j$ after propagating $p$ periods is
\begin{subequations}
\begin{align}
\label{eq:propsum}
\tilde{a}_j(p)=&\sum_s{\phi_{js}\mu_{1s}^pw_s(0)}, \\
\label{eq:propmat}
{\rm i.e. }~\mathbf{\tilde{a}}(p)=&\mathbf{\phi}\,\mathbf{L}^p \mathbf{w}(0),
\end{align}
\end{subequations}
where $\mathbf{L}=\text{diag}(\mu_{1s})$.
Writing the initial condition $\mathbf{w}(0)$ in terms of waveguide rather than supermode amplitudes, we see that the field at any integer number of periods $p$ along the PCW array is
\begin{equation}
	\mathbf{\tilde a}(p) = \mathbf{\phi}\, \mathbf{L}\, \mathbf{\phi}^{-1} \mathbf{\tilde a}(0),
\end{equation}
where $\mathbf{\phi} = \mathbf{\phi}^{-1}$.

Therefore, knowing only the dispersion relation, we can analytically calculate the amplitudes $\mathbf{\tilde a}(p)$ of each waveguide mode at an arbitrary number of periods $p$ into the structure.
The actual field in each waveguide may be calculated from $\mathbf{\tilde a}(p)$, and the field $H_{{\rm swg}}(\mathbf{r})$ of the single waveguide mode over the domain $0 < x < d$.
Recall that in Eq.~\eqref{eq:tilde_transformation}, we transformed the Bloch mode amplitudes $\mathbf{c}$ into a mathematically convenient basis and wrote them as $\mathbf{\tilde c}$.
Unlike in Sec.~\ref{singlewgsec}, we cannot set the origin of the coordinate system such that $a_j = \tilde a_j$ for all waveguides $j$, therefore we must explicitly perform a transformation into a more natural basis.
We map each $\tilde a_j$ to the amplitude $a_j$ of the single mode waveguide in the unit cell $p$ periods into the structure, by writing
\begin{equation}
	a_j = \kappa_j \tilde a_j,
\end{equation}
where $\kappa_j = \exp(ik_x (x_R-pd))$, with $x_R$ the coordinate of the unit cell in waveguide $j$ (more specifically, the point in that unit cell marked with a cross in Fig.~\ref{scat1}(a), $p$ periods into the structure.
The origin of coordinates may be chosen such that $x_R = pd$ for every unit cell $p$ along waveguide $j=1$: then $\kappa_j = 1$ for all waveguides that are in line with waveguide 1, and $\kappa_j = exp(ik_xd/2)$ for waveguides staggered by $\Delta x = d/2$ with respect to waveguide 1.
Here, $k_x$ is approximated to that for the single waveguide mode.

Finally, we write the field in the waveguide array as a superposition of translated single waveguide modes $H_{{\rm swg}}(\mathbf{r-R}_j)$ and their amplitudes $\mathbf{\tilde a}(p)$:
\begin{equation}
	H(\mathbf{r}) = \sum_{j} \kappa_j \tilde a_j H_{{\rm swg}}(\mathbf{r}-\mathbf{R}_j),
\end{equation}
where $\mathbf{R}_j = (x_R, y_R)$ is the coordinates of the nearest unit cell to $\mathbf{r}$ inside waveguide $j$.

If we neglect the evanescent tails of single waveguide modes, then we can write the field in or near waveguide $j$ of the array in terms of one waveguide mode only:
\begin{equation}
	H_j(\mathbf{r}) = \kappa_j \tilde a_j H_{{\rm swg}}(\mathbf{r}-\mathbf{R}_j).
\end{equation}

If we confine our interest to the field inside the waveguide at a single point per unit cell, then $\mathbf{r} - \mathbf{R}_j$ is constant and only one value $H_{{\rm swg}}(\mathbf{r}-\mathbf{R}_j)$ is used; the mode may be renormalized such that $H_{{\rm swg}}(\mathbf{r}-\mathbf{R}_j) = 1$.
Then the field throughout the waveguide array is given by the elements $a_j$ of $\mathbf{a}(p)$, which can be calculated analytically.

\section{Results}
\label{results}
We now apply the methods we have developed to calculate dispersion relations for arrays of coupled waveguides, as well as the propagation of light through them.
Where the use of conventional numerical tools is practical, we compare the results of our method to these.
For simplicity, in all our examples we consider arrays constructed from the same two PCs.
Both PCs are regular hexagonal lattices of cylindrical inclusions with radius $r=0.3~d$, in a dielectric background with $n = 3$.
The difference between the PCs is that the barrier, PC~1, has air holes with $n=1$, whereas the waveguide material, PC~2, has inclusions with $n=1.5$.
Our method also applies to more conventional W1 PC waveguides, and can easily be extended to dispersion engineered waveguides: we consider an example with infiltrated holes in PC~2 in order to demonstrate that our method works even when the waveguiding region is not a uniform dielectric.

We consider two arrays based on this waveguide.
The first is a system of $M = 4$ waveguides, each separated by a barrier of $\ell = 4$ rows of PC~1.
The second array is a system of $M=31$ waveguides, also separated by $\ell = 4$ rows of PC~1.
The second array involves a structure with a width of over 150 unit cells, which is not feasible to simulate using conventional numerical tools.
The separation of $\ell = 4$ means that the waveguides are staggered.

We consider 5 Bloch modes in each PC, which means that for each frequency-$k_x$ pair, testing for the existence of a supermode only involves calculating the determinants of $M=4$ different $5 \times 5$ matrices (Eq.~\eqref{virtualequations}).
At the frequencies we consider, we could get results of acceptable accuracy using only 2 Bloch modes.

In Sec.~\ref{results1}, we use Eq.~\eqref{virtualequations} to calculate the dispersion relations of these structures.
Then, in Sec.~\ref{results2} we calculate the discrete diffraction pattern of the field as it propagates through the waveguide array.

\subsection{PCW Array Dispersion relations}
\label{results1}
\begin{figure}[ht]
\centering
\includegraphics[width=8cm]{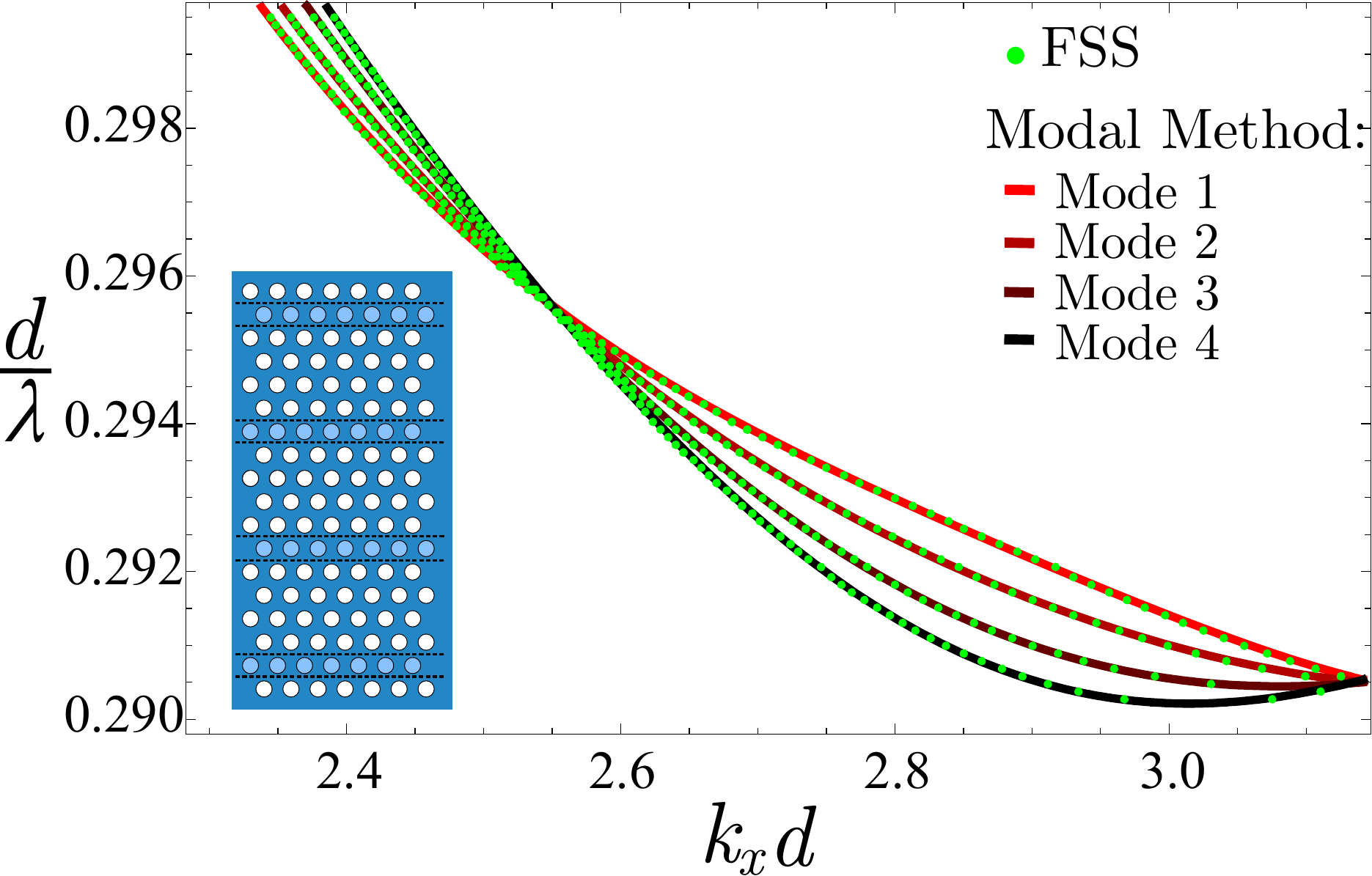}
\caption{Comparison of the dispersion relation calculated using the FSS (green dots), and our semi-analytic modal method (red lines) for a four waveguide array with waveguides separated by four rows of inclusions (shown in inset). Background refractive index, $n=3$, cylindrical inclusions of index $n=1$ with the waveguide created by changing the refractive index of the waveguide inclusions to $n=1.5$.}\label{formresults}
\end{figure}

First, we calculate the dispersion curves of the supermodes of the 4 waveguide system described above, for light polarized with the $H$ field out of the plane, at normalized frequencies $d/\lambda \in (0.29,0.30)$.
The dispersion curves are shown in Fig.~\ref{formresults}.
At most frequencies there are four supermodes based on the even single waveguide mode, and their dispersion curves agree well with those calculated using the Fictitious Source Superposition (FSS) method, which is considered to be highly accurate for this kind of problem \cite{botten2006highly}.

\begin{figure}[hb]
\centering
\includegraphics[width=8cm]{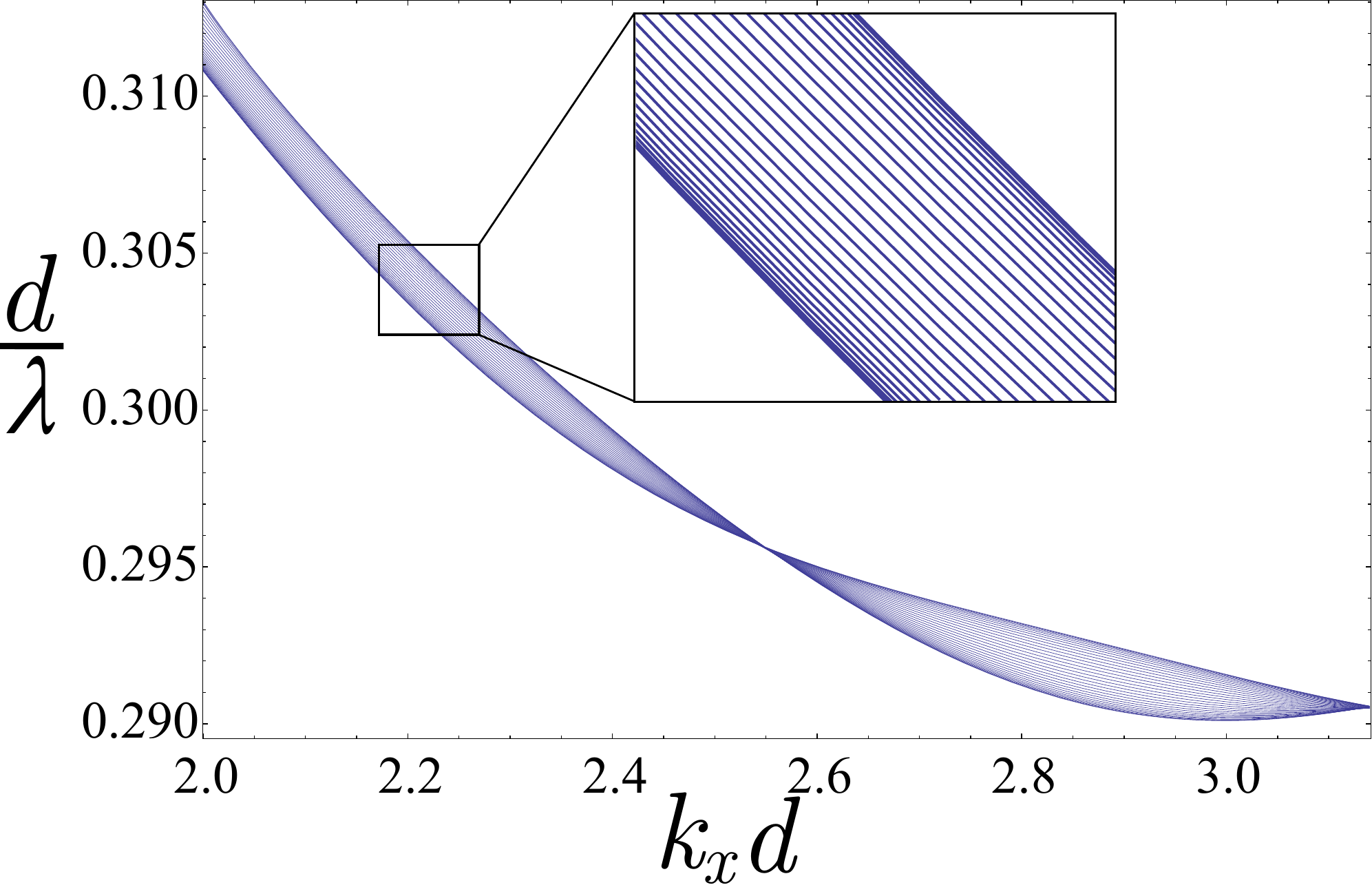}
\caption{The dispersion curve for a $M=31$ PCW array with four rows of inclusions between each waveguide. The background index is $n=3$, the bulk photonic crystal inclusions have index $n=1$, and the waveguide is formed with inclusions of index $n=1.5$.}\label{thirtydispss}
\end{figure}

Having established the accuracy of our method with a relatively small array of 4 waveguides, we now show its power by finding the dispersion curves of the supermodes of a 31-waveguide array (Fig. \ref{thirtydispss}).
We see that for this increased number of waveguides and supermodes, the dispersion curves of the supermodes start to form a band, but the dispersion curve still exhibits the braiding behavior shown for the coupled PCW system \cite{brownless2010coupled}.

\subsection{Driven PCW Arrays}
\label{results2}
\begin{figure}[hb]
\centering
\includegraphics[width=8cm]{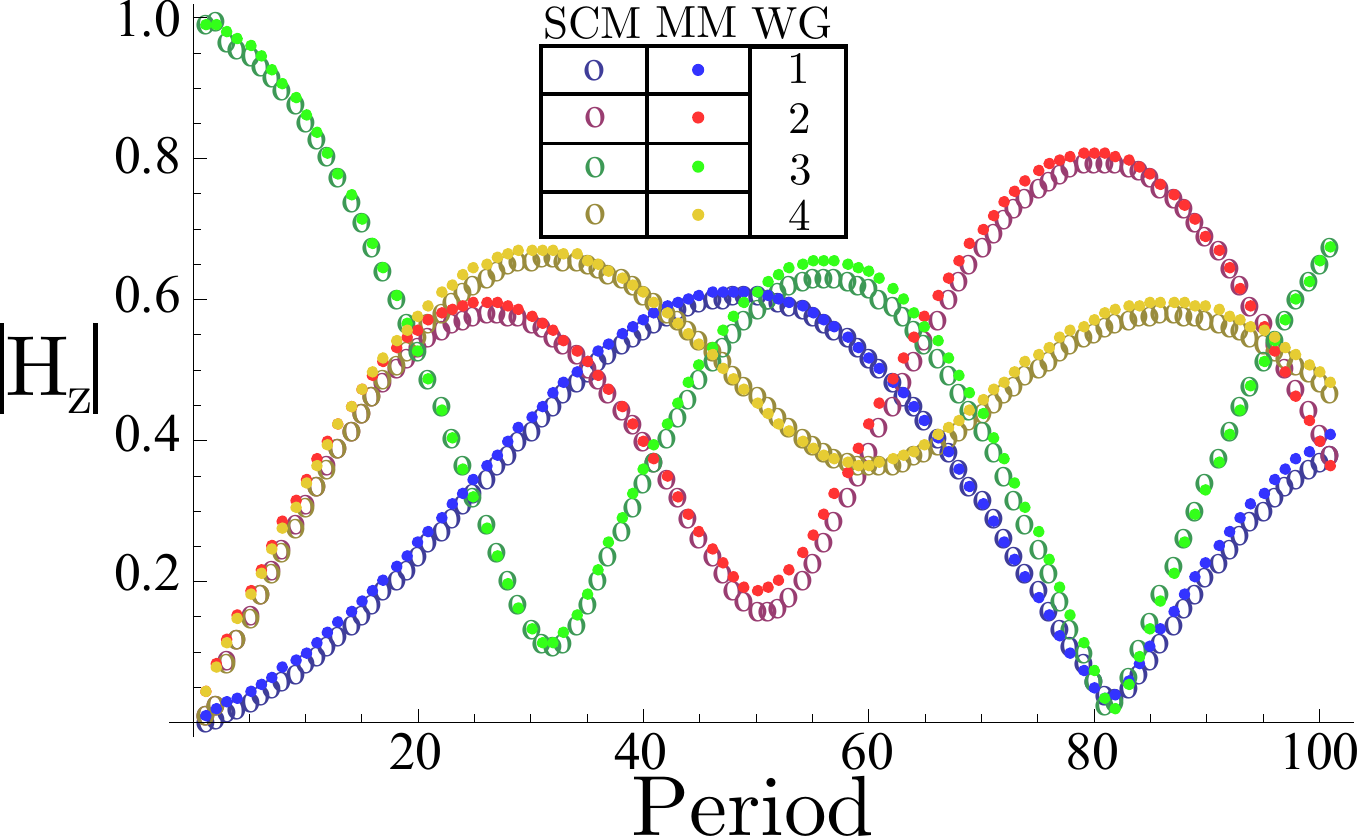}
\caption{A comparison of methods for calculating the field throughout the waveguide array at frequency $d/\lambda = 0.293$. One point is taken per unit cell of the array shown in Fig.~\ref{formresults}. The colors represent the different waveguides; the open markers are from Bloch mode supercell calculations (SCM) and the solid markers are results from our modal method (MM).}\label{comparison}
\end{figure}

We now turn our attention to field propagation through the two PCW arrays.
We seek to simulate a waveguide array where the field is incident from a single waveguide.
In the $M = 4$ waveguide array, we set $\mathbf{\tilde a}(0) = (0,0,1,0)^T$, corresponding to an initial condition where the entire field is in the 3rd waveguide.
At a normalized frequency of $d/\lambda = 0.293$, there are four propagating supermodes, at $k_x d = -2.68, -2.70, -2.74,$ and $-2.80$.
A single waveguide with these parameters has $k_x d = -2.72$
We choose the negative values of $k_x$ as they are associated with positive group velocity (Fig. \ref{formresults}).
We calculate their field at one point per unit cell over the first 100 periods of the waveguide array, i.e., we calculate $\mathbf{\tilde a}(p)$ from $p=0$ to $p=100$ (Fig. \ref{comparison}).
The approximation made when writing $\mathbf{\tilde x}_s \simeq \mathbf{\tilde e}$ has normalized residual errors of $(2.2, 0.3, 0.3, 2.1)\times 10^{-3}$ for the four supermodes.

These results are indistinguishable from the results of an FSS simulation, although neither method simulates the coupling at the interface between the input waveguide and the waveguide array.
In order to include coupling effects, we use a well established Bloch modal supercell scattering method \cite{botten2004bloch} to model propagation through a single waveguide into the waveguide array.  We sample the field at the same points in each unit cell for each waveguide in the array.
There is good agreement between the supercell method and our results (Fig. \ref{comparison}).

\begin{figure*}[thb]
\centering
\includegraphics[width=12cm]{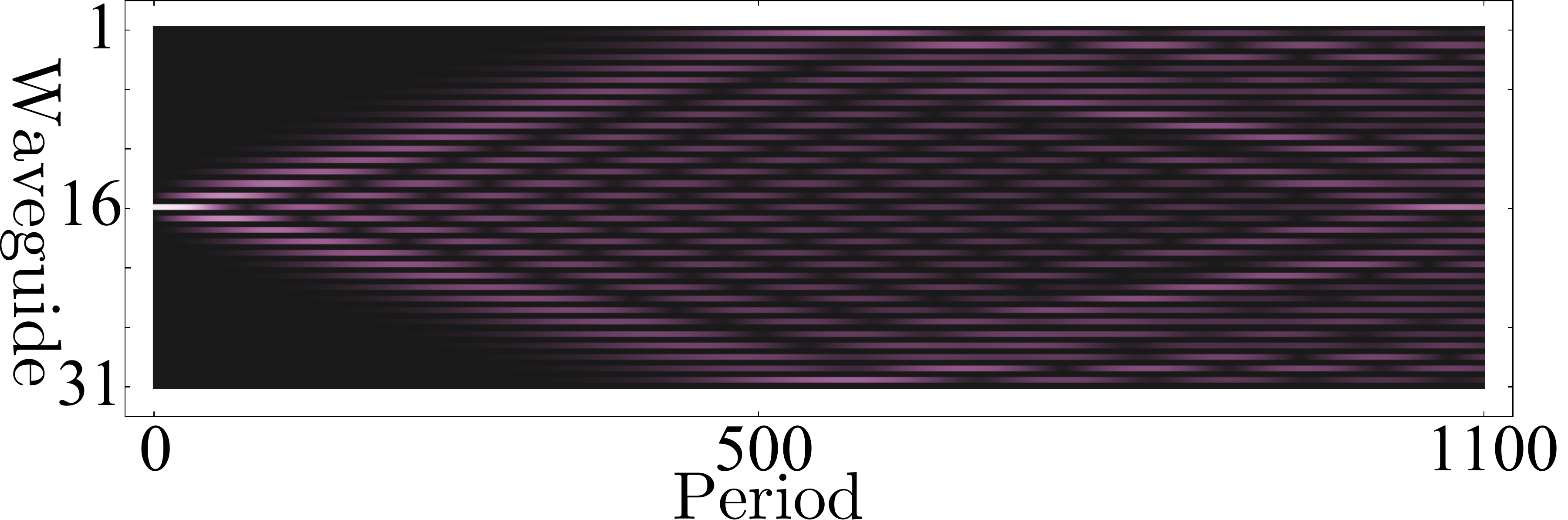}
\caption{Propagation of light through the $M=31$ PCW array described in Fig.~\ref{thirtydispss}. The initial condition is that all the energy begins in the PCW mode centered at waveguide $j=16$ with a frequency of $d/\lambda=0.293$.}\label{thirtyprop}
\end{figure*}
Now we consider the system of $M = 31$ coupled waveguides.
Again, we start with all the field in the central waveguide, so all elements ${\tilde a}_j$ of $\mathbf{\tilde a}(0)$ are 0, except ${\tilde a}_{16} = 1$.
We calculate the propagation of the field through 1100 periods of the waveguide array (Fig. \ref{thirtyprop}).
For the first 450 periods, a typical discrete diffraction pattern is observed.
After around 500 periods, the diffracted light reaches the outer waveguides, and edge effects start to play a significant role in the field profile.
Once the $k_x$ values of the supermodes are calculated, generation of the data for Fig. \ref{thirtyprop} takes under 10 seconds on a desktop computer.

\section{Discussion and Conclusion}
\label{conclusion}
The semi-analytical formulation developed here provides an accurate modal method to solve for the modes of coupled PCW waveguides. An advantage of this method is that the computational time increases linearly as the number of waveguides increases, in the sense that for $M$ waveguides, $M$ equations of the form of (\ref{finalcondition}) need to be solved to obtain the PCW modes at each frequency. The advantage here is that each mode is identified by its symmetry and solved for independently. Therefore, the memory requirements remain constant as $M$ increases. This contrasts with supercell methods, where as the number of waveguides increases, the computation domain grows, increasing both the required memory and computation time. Furthermore, in supercell methods the modes of different symmetry are not readily identifiable, making crossings in the dispersion curves, such as those shown in Figures \ref{formresults} and \ref{thirtydispss} difficult to recognize.

Our method is, in essence, a tight-binding formulation and therefore it warrants comparison to a conventional tight-binding formulation based on overlap integrals  \cite{yeh1988optical}. In such conventional formalisms one solves for the isolated single PCW mode and assumes that the modes of the coupled PCWs can be written in terms of a superposition of isolated single PCW modes. The interaction between adjacent PCWs is found by  computing overlap integrals. The formulation presented here is identical in terms of the assumptions it is based on, i.e. adjacent waveguides couple weakly, however in place of overlap integrals interaction between waveguides is modeled by a transmission matrix. There is one key advantage in using our modal method based calculation: we find the PCW modes by finding the propagation constants $k_x$ satisfying (\ref{finalcondition}) at a given frequency. In contrast, an overlap integral based formulation for PCWs gives the shift in frequency of supermodes from the single PCW mode. The former is much more desirable for modeling propagation problems, as once an operation frequency is chosen the relevant modes can be obtained in a single set of calculations.
At the level of Bloch modes and supermodes, the similarities between the acoustic and electromagnetic treatments mean that this modal method may also be applied to arrays of elastic waveguides \cite{adams2009guided, sun2005analyses}.

In the work of de Sterke {\it et al} \cite{de2004modes}, it is shown that although the fundamental mode can be either an even or odd superposition of PCW modes, when taking a cross section along the PCWs, the fundamental mode always possesses one less node than the second mode.  Here we find that, when taking a cross section along the coupled PCWs, node counting cannot be used to predict the order of modes in the hexagonal lattice. This is because the oscillation theorem only holds for one dimensional Sturm-Liouville problems. In the case of square lattice PCWs, the modes have a frequency below the first Wood anomaly \cite{de2004modes}. This allows for the use of a scalar approach where coupled PCWs are mapped on to coupled uniform waveguides. The rich physics found in hexagonal lattice PCW arrays, is due to the fact that there are two dominant Bloch modes and a vectorial approach is needed. We have observed that near the center of the Brillouin zone, where one operates below the first Wood anomaly, the oscillation theorem holds in the hexagonal lattice. However, as the Bloch wavevector approaches the Brillouin zone edge, crossing the first Wood anomaly and especially upon entering the braided region, the modes become inherently vectorial in nature, ie. their Bloch mode expansion requires at least two Bloch modes, and thus cannot be mapped on to a scalar amplitude like in the square lattice. Therefore there is no general node counting algorithm governing the order of modes in hexagonal PCW arrays.

\section*{Acknowledgments}
This research was conducted by the Australian Research Council Centre
of Excellence for Ultrahigh Bandwidth Devices for Optical Systems
(project number~CE110001018).


\end{document}